\documentclass[a4paper,UKenglish]{lipics-v2016}
 
\usepackage{microtype}

\usepackage{graphicx}
\usepackage{amsmath,amssymb}
\newcommand{\REV}[1]{\ensuremath{\overline{#1}}}
\newcommand{\RLBWT}{\ensuremath{\mathsf{RLBWT}}}
\newcommand{\RLCSA}{\ensuremath{\mathsf{RLCSA}}}
\newcommand{\ST}{\ensuremath{\mathsf{ST}}}

\newcommand{\SA}{\ensuremath{\mathsf{SA}}}

\newcommand{\BWT}{\ensuremath{\mathsf{BWT}}}
\newcommand{\CDAWG}{\ensuremath{\mathsf{CDAWG}}}
\newcommand{\SSA}{\ensuremath{\mathsf{SSA}}}
\newcommand\SP[1]{\mathtt{sp}(#1)}
\newcommand\EP[1]{\mathtt{ep}(#1)} 
\newcommand\INTERVAL[1]{\mathtt{range}(#1)}
\newcommand{\INTERVALFUNCTION}{\ensuremath{\mathbb{I}}}

\newcommand{\occ}
  {\ensuremath{\mathsf{occ}}}
\newcommand{\runs}{r}

\title{Practical combinations of repetition-aware data structures\footnote{This work was partially supported by Academy of Finland under grant 284598 (Center of Excellence in Cancer Genetics Research).}}

\author[1]{Djamal Belazzougui}
\author[2]{Fabio Cunial}
\author[3,4]{Travis Gagie}
\author[5]{Nicola Prezza}
\author[6]{Mathieu Raffinot}

\affil[1]{CERIST, Algeria. 
}
\affil[2]{Max Planck Institute of Molecular Cell Biology and Genetics, Dresden, Germany.
}
\affil[3]{Department of Computer Science, University of Helsinki, Finland.}
\affil[4]{Helsinki Institute for Information Technology, Finland.}
\affil[5]{Department of Mathematics and Computer Science, University of Udine, Italy.}
\affil[6]{Laboratoire Bordelais de Recherche en Informatique, CNRS, Bordeaux, France.}

\authorrunning{D. Belazzougui et al.} 

\Copyright{D. Belazzougui et al.}

\subjclass{E.1 Data structures, F.2.2 Pattern matching.}
\keywords{repetitive strings, locate, count, run-length encoded BWT, Lempel-Ziv factorization, CDAWG.}

\EventEditors{John Q. Open and Joan R. Acces}
\EventNoEds{2}
\EventLongTitle{42nd Conference on Very Important Topics (CVIT 2016)}
\EventShortTitle{CVIT 2016}
\EventAcronym{CVIT}
\EventYear{2016}
\EventDate{December 24--27, 2016}
\EventLocation{Little Whinging, United Kingdom}
\EventLogo{}
\SeriesVolume{42}
\ArticleNo{23}

\begin{document}

\maketitle

\begin{abstract}
Highly-repetitive collections of strings are increasingly being amassed by genome sequencing and genetic variation experiments, as well as by storing all versions of human-generated files, like webpages and source code. Existing indexes for locating all the exact occurrences of a pattern in a highly-repetitive string take advantage of a single measure of repetition. However, multiple, distinct measures of repetition all grow sublinearly in the length of a highly-repetitive string. In this paper we explore the practical advantages of combining data structures whose size depends on distinct measures of repetition. 
The main ingredient of our structures is the run-length encoded BWT (RLBWT), which takes space proportional to the number of runs in the Burrows-Wheeler transform of a string. We describe a range of practical variants that combine RLBWT with the set of boundaries of the Lempel-Ziv 77 factors of a string, which take space proportional to the number of factors. Such variants use, respectively, the RLBWT of a string and the RLBWT of its reverse, or just one RLBWT inside a bidirectional index, 
or just one RLBWT with support for unidirectional extraction. We also study the practical advantages of combining RLBWT with the compact directed acyclic word graph of a string, a data structure that takes space proportional to the number of one-character extensions of maximal repeats. 
Our approaches are easy to implement, and provide competitive tradeoffs on significant datasets.
\end{abstract}

\section{Introduction}\label{sec:introduction}

Locating all the exact occurrences of a string in a massive, highly-repetitive collection of similar texts is a fundamental primitive in the post-genome era, in which genomes from multiple related species, from multiple strains of the same species, and from multiple individuals, are being sequenced at an increasing pace. Most data structures designed for such repetitive collections take space proportional to a specific measure of repetition, for example the number $z$ of factors in a Lempel-Ziv parsing \cite{arroyuelo2012stronger,kreft2013compressing}, or the number $r$ of runs in a Burrows-Wheeler transform \cite{MakinenNSV10}. 
In previous work we showed how to achieve competitive theoretical tradeoffs between space and time in locate queries, by combining data structures that depend on distinct measures of repetition, where all such measures grow sublinearly in the length of a highly-repetitive string \cite{belazzougui2015composite}. Specifically, we described a data structure that takes approximately $O(z+r)$ words of space, and that reports all the occurrences of a pattern of length $m$ in a string of length $n$ in $O(m(\log{\log{n}} + \log{z}) + \mathtt{pocc} \cdot \log^{\epsilon}{z} + \mathtt{socc} \cdot \log{\log{n}})$ time, where $\mathtt{pocc}$ and $\mathtt{socc}$ are the number of primary and of secondary occurrences, respectively (see Section \ref{sec:stringIndexes}). This compares favorably to the $O(m^{2}h+(m+\mathtt{occ})\log{z})$ reporting time of Lempel-Ziv 77 (LZ77) indexes \cite{kreft2013compressing}, where $h$ is the height of the parse tree. It also compares favorably in space to solutions based on the run-length encoded BWT (RLBWT) and on suffix array samples
\cite{MakinenNSV10}, which take $O(n/k + r)$ words of space to achieve
$O(m\log{\log{n}} + k \cdot \mathtt{occ} \cdot \log{\log{n}})$ reporting
time, where $k$ is the sampling rate. 
We also introduced a data structure whose size depends on the number of right-extensions of maximal repeats, and that reports all the $\mathtt{occ}$ occurrences of a pattern in $O(m\log{\log{n}} + \mathtt{occ})$ time. The main component of our constructions is the RLBWT, which we use for counting the number of occurrences of a pattern, and which we combine with the CDAWG and with data structures from Lempel-Ziv indexes, rather than with suffix array samples, for answering locate queries.

%

In this paper we engineer a range of practical variants of such approaches, and we compare their space-time tradeoffs to a representative set of state-of-the-art indexes for repetitive collections, including RLCSA \cite{MakinenNSV10}, a number of LZ77 implementations \cite{kreft2010self}, and a recent implementation of the hybrid index \cite{valenzuela2016chico}. One of our indexes based on RLBWT and LZ77 factors uses an amount of memory comparable to LZ77 indexes, but it answers count queries between two and four orders of magnitude faster than all LZ77 and hybrid index implementations. For long patterns, our index uses less space than the hybrid index, and it answers locate queries between one and two orders of magnitude faster than a number of LZ77 implementations, and as fast as the fastest LZ77 implementation. With short patterns, our index based on RLBWT and CDAWG answers locate queries between four and ten times faster than a version of RLCSA that uses comparable memory, and with extremely short patterns our index achieves speedups even greater than ten with respect to RLCSA.

\section{Preliminaries}

\subsection{Strings} \label{sec:strings}

Let $\Sigma = [1..\sigma]$ be an integer alphabet, let $\# = 0 \notin \Sigma$ be a separator, and let $T \in [1..\sigma]^{n-1}$ be a string. We denote by $\overline{T}$ the reverse of $T$, and by $\mathcal{P}_{T\#}(W)$ the set of all starting positions of a string $W \in [0..\sigma]^+$ in the circular version of $T\#$. We set $\Sigma^{r}_{T\#}(W)=\{a \in [0..\sigma] : |\mathcal{P}_{T\#}(Wa)|>0 \}$ and  $\Sigma^{\ell}_{T\#}(W)=\{a \in [0..\sigma] : |\mathcal{P}_{T\#}(aW)|>0 \}$. A \emph{repeat} $W \in \Sigma^+$ is a string that satisfies $|\mathcal{P}_{T\#}(W)|>1$. A repeat $W$ is \emph{right-maximal} (respectively, \emph{left-maximal}) iff $|\Sigma^{r}_{T\#}(W)|>1$ (respectively, iff $|\Sigma^{\ell}_{T\#}(W)|>1$). A \emph{maximal repeat} is a repeat that is both left- and right-maximal. We say that a maximal repeat $W$ is \emph{rightmost} (respectively, \emph{leftmost}) if no string $WV$ with $V \in [0..\sigma]^+$ is left-maximal (respectively, if no string $VW$ with $V \in [0..\sigma]^+$ is right-maximal).

For reasons of space we assume the reader to be familiar with the notion of \emph{suffix tree} $\ST_{T\#} = (V,E)$ and of \emph{suffix-link tree} of $T\#$, which we do not define here. We denote by $\ell(\gamma)$, or equivalently by $\ell(u,v)$, the label of edge $\gamma=(u,v) \in E$, and we denote by $\ell(v)$ the string label of node $v \in V$. It is well known that a string $W$ is right-maximal (respectively, left-maximal) in $T\#$ iff $W=\ell(v)$ for some internal node $v$ of $\ST_{T\#}$ (respectively, iff $W=\REV{\ell(v)}$ for some internal node $v$ of $\ST_{\REV{T}\#}$). Since left-maximality is closed under prefix operation, there is a bijection between the set of all maximal repeats of $T\#$ and the set of all nodes of the suffix tree of $T\#$ that lie on paths that start from the root and that end at nodes labelled by rightmost maximal repeats. Symmetrically, since right-maximality is closed under suffix operation, there is a bijection between the set of all maximal repeats and the set of all nodes of the suffix tree of $\REV{T}\#$ that lie on paths that start from the root and that end at nodes labelled by leftmost maximal repeats.

The \emph{compact directed acyclic word graph} of $T\#$ (denoted by $\CDAWG_{T\#}$ in what follows) is the minimal compact automaton that recognizes the set of suffixes of $T\#$ \cite{blumer1987complete,CrochemoreV97}. It can be seen as the minimization of $\ST_{T\#}$ in which all leaves are merged to the same node (the sink) that represents $T\#$ itself, and in which all nodes except the sink are in one-to-one correspondence with the maximal repeats of $T\#$ \cite{Raffinot2001} (the source corresponds to the empty string). The set of accepting nodes consists of the sink and of all maximal repeats that also occur as a suffix of $T\#$. Like in the suffix tree, transitions are labelled by substrings of $T\#$. Since a maximal repeat corresponds to a subset of its right-maximal suffixes, $\CDAWG_{T\#}$ can be built by putting in the same equivalence class all nodes of $\ST_{T\#}$ that belong to the same maximal unary path of explicit Weiner links. Note also that the subgraph of $\ST_{T\#}$ induced by maximal repeats is isomorphic to a spanning tree of $\CDAWG_{T\#}$.

For reasons of space we assume the reader to be familiar with the notion and uses of the Burrows-Wheeler transform of $T$, including the $C$ array, LF mapping, and backward search. In this paper we use $\BWT_{T\#}$ to denote the BWT of $T\#$, and we use $\INTERVAL{W} = [\SP{W}..\EP{W}]$ to denote the lexicographic interval of a string $W$ in a BWT that is implicit from the context. We say that $\BWT_{T\#}[i..j]$ is a \emph{run} iff $\BWT_{T\#}[k]=c \in [0..\sigma]$ for all $k \in [i..j]$, and moreover if any substring $\BWT_{T\#}[i'..j']$ such that $i' \leq i$, $j' \geq j$, and either $i' \neq i$ or $j' \neq j$, contains at least two distinct characters. It is well known that repetitions in $T\#$ induce runs in $\BWT_{T\#}$: for example, the BWT of $W^{k}\#$ consists of $r\leq |W|$ runs of length at least $k$ and of a run of length one. We denote by $\runs_{T\#}$ the number of runs in $\BWT_{T\#}$, and we call \emph{run-length encoded BWT} (denoted by $\RLBWT_{T\#}$) any representation of $\BWT_{T\#}$ that takes $O(\runs_{T\#})$ words of space, and that supports rank and select operations (see e.g. \cite{makinen2005succinct1,MakinenNSV10,SirenVMN08}). 
Since the difference between $\runs_{T\#}$ and $\runs_{\REV{T}\#}$ is negligible in practice, to simplify notation we denote both of them by $\runs$ when $T$ is implicit from the context.

The \emph{Lempel-Ziv 77 factorization} of $T$ \cite{ziv1977universal}, abbreviated with LZ77 in the rest of the paper, is the greedy decomposition $T_1 T_2 \cdots T_{z}$ of $T$ defined as follows. Assume that $T$ is virtually preceded by the set of distinct characters in its alphabet, and assume that $T_1 T_2 \cdots T_i$ has already been computed for some prefix of length $k$ of $T$: then, $T_{i+1}$ is the longest prefix of $T[k+1..n]$ such that there is a $j \leq k$ that satisfies $T[j..j+|T_{i+1}|-1] = T_{i+1}$. 

In the rest of the paper we drop subscripts whenever they are clear from the context.

\subsection{String indexes} \label{sec:stringIndexes}

The \emph{run-length compressed suffix array} of $T\#$, denoted by $\RLCSA_{T\#}$ in what follows, consists of a run-length compressed rank data structure for $\BWT_{T\#}$, and of a sampled suffix array, denoted by $\SSA_{T\#}$ \cite{MakinenNSV10}. Given a pattern $P \in [1..\sigma]^m$, we use the rank data structure to find the interval of $\BWT_{T\#}$ that contains all characters that precede the occurrences of $P$ in $T\#$: the length of this interval, uncompressed, is the number of such occurrences. To locate a specific occurrence, we start at the character that precedes it in $\BWT_{T\#}$ and we use rank queries to move backward, until we reach a character whose position has been sampled. Thus, the average time for locating an occurrence is inversely proportional to the size of $\SSA_{T\#}$, and fast locating needs a large SSA regardless of the compressibility of the dataset. 
M\"akinen et al.\ suggested ways to reduce the size of the SSA \cite{MakinenNSV10}, but they did not perform well enough in real repetitive datasets for the authors to include them in the software they released.

For reasons of space we assume the reader to be familiar with LZ77 indexes (see e.g. \cite{gagie2014lz77,karkkainen1996lempel}). Here we just recall that a \emph{primary occurrence} of a pattern $P$ in $T$ is one that crosses or ends at a phrase boundary in the LZ77 factorization $T_1 T_2 \cdots T_{z}$ of $T$. All other occurrences are called \emph{secondary}. Once we have computed primary occurrences, locating all $\mathtt{socc}$ secondary occurrences reduces to two-sided range reporting, and it takes $O(\mathtt{socc} \cdot \log{\log{n}})$ time with a data structure of $O(z)$ words of space \cite{karkkainen1996lempel}. To locate primary occurrences, we use a data structure for four-sided range reporting on a \(z \times z\) grid, with a marker at \((x, y)\) if the $x$-th LZ factor in lexicographic order is preceded in the text by the lexicographically $y$-th reversed prefix ending at a phrase boundary. This data structure takes $O(z)$ words of space, and it returns all the phrase boundaries that are immediately followed by a factor in the specified range, and immediately preceded by a reversed prefix in the specified range, in $O((1+k)\log^{\epsilon}{z})$ time, where $k$ is the number of phrase boundaries reported \cite{chan2011orthogonal}. K\"arkk\"ainen and Ukkonen used two PATRICIA trees \cite{morrison1968patricia}, one for the factors and the other for the reversed prefixes ending at phrase boundaries \cite{karkkainen1996lempel}. 
To locate primary occurrences, we query the first tree for the range of distinct factors in left-to-right lexicographic order that start with $P[i+1..m]$, and we query the second tree for the range of reversed prefixes $\REV{T[1..p_i-1]}$ starting with $\REV{P[1..i]}$, for all $i \in [1..m]$. The ranges returned by the trees are correct iff any factor starts with $P[i+1..m]$ and any reversed prefix at a phrase boundary starts with $\REV{P[1..i]}$. To check the first range, we choose any factor in the range and compare its first $m-i$ characters to $P[i+1..m]$. To check the second range, we choose any reversed prefix in the range and compare its first $i$ characters to $\REV{P[1..i]}$. This takes $O(m)$ time for every $i$, thus $O(m^2)$ time in total, assuming that $T$ is not compressed. Replacing the uncompressed text by an augmented compressed representation, we can store $T$ in $O(z \log{n})$ space such that later, given $P$, we can find all $\occ$ occurrences of $P$ in $O(m\log{m} + \occ \cdot \log{\log{n}})$ time~\cite{gagie2014lz77}.

If we know in advance that all patterns will be of length at most $M$, then we can store in a FM-index the substrings of $T$ consisting of characters within distance $M$ of the nearest phrase boundary, and use that to find primary occurrences. This approach, called \emph{hybrid indexing}, has been proposed several times recently: see e.g. \cite{valenzuela2016chico} and references therein for more details.

\subsection{Composite repetition-aware string indexes} \label{sec:compositeStringIndexes}

It is possible to combine $\RLBWT_{T\#}$ with the set of all starting positions $p_1,p_2,\dots,p_z$ of LZ factors of $T$, building a data structure that takes $O(z+r)$ words of space, and that reports all the $\mathtt{pocc}$ primary occurrences of a pattern $P \in [1..\sigma]^m$ in $O(m(\log{\log{n}}+\log{z}) + \mathtt{pocc} \cdot \log^{\epsilon}{z})$ time \cite{belazzougui2015composite}. Since such data structure is at the core of the paper, we summarize how it works in what follows.

The same primary occurrence of $P$ in $T$ can cover up to $m$ boundaries between two LZ factors. Thus, we consider every possible way of placing, inside $P$, the rightmost boundary between two factors, i.e. every possible split of $P$ in two parts $P[1..k-1]$ and $P[k..m]$ for $k \in [2..m]$, such that $P[k..m]$ is either a factor or a proper prefix of a factor. For every such $k$, we use four-sided range reporting queries to list all the occurrences of $P$ in $T$ that conform to the split, as described in Section \ref{sec:stringIndexes}. 
We encode the sequence $p_1,p_2,\dots,p_z$ implicitly, as follows: we use a bitvector $\mathtt{last}[1..n]$ such that $\mathtt{last}[i]=1$ iff $\SA_{\REV{T}\#}[i]=n-p_j+2$ for some $j \in [1..z]$, i.e. iff $\SA_{\REV{T}\#}[i]$ is the last position of a factor. We represent such bitvector as a predecessor data structure with partial ranks, using $O(z)$ words of space \cite{Wi83}. Let $\ST_{T\#} = (V,E)$ be the suffix tree of $T\#$, and let $V'=\{v_1,v_2,\dots,v_z\} \subseteq V$ be the set of loci in $\ST_{T\#}$ of all LZ factors of $T$. Consider the list of node labels $L = \ell(v_1),\ell(v_2),\dots,\ell(v_z)$, sorted in lexicographic order. It is easy to build a data structure that takes $O(z)$ words of space, and that implements in $O(\log{z})$ time function $\INTERVALFUNCTION(W,V')$, which returns the (possibly empty) interval of $W$ in $L$ (see e.g. \cite{belazzougui2015composite}). Together with $\mathtt{last}$, $\RLBWT_{T\#}$ and $\RLBWT_{\REV{T}\#}$, this data structure is the output of our construction.

Given $P$, we first perform a backward search in $\RLBWT_{T\#}$ to determine the number of occurrences of $P$ in $T\#$: if this number is zero, we stop. During backward search, we store in a table the interval $[i_k..j_k]$ of $P[k..m]$ in $\BWT_{T\#}$ for every $k \in [2..m]$. Then, we compute the interval $[i'_{k-1}..j'_{k-1}]$ of $\REV{P[1..k-1]}$ in $\BWT_{\REV{T}\#}$ for every $k \in [2..m]$, using backward search in $\RLBWT_{\REV{T}\#}$: if $\mathtt{rank}_{1}(\mathtt{last},j'_{k-1}) - \mathtt{rank}_{1}(\mathtt{last},i'_{k-1}-1) = 0$, then $P[1..k-1]$ never ends at the last position of a factor, and we can discard this value of $k$. Otherwise, we convert $[i'_{k-1}..j'_{k-1}]$ to the interval $[\mathtt{rank}_{1}(\mathtt{last},i'_{k-1}-1)+1 .. \mathtt{rank}_{1}(\mathtt{last},j'_{k-1})]$ of all the reversed prefixes of $T$ that end at the last position of a factor. Rank operations on $\mathtt{last}$ can be implemented in $O(\log{\log{n}})$ time using predecessor queries. We get the lexicographic interval of $P[k..m]$ in the list of all distinct factors of $T$ using operation $\INTERVALFUNCTION(P[k..m],V')$, in $O(\log z)$ time. We use such intervals to query the four-sided range reporting data structure.

It is also possible to combine $\RLBWT_{T\#}$ with $\CDAWG_{T\#}$, building a data structure that takes $O(e_{T\#})$ words of space, and that reports all the $\mathtt{occ}$ occurrences of $P$ in $O(m\log{\log{n}} + \mathtt{occ})$ time, where $e_{T\#}$ is the number of right-extensions of maximal repeats of $T\#$ \cite{belazzougui2015composite}. Specifically, for every node $v$ in the CDAWG, we store $|\ell(v)|$ in a variable $v.\mathtt{length}$. Recall that an arc $(v,w)$ in the CDAWG means that maximal repeat $\ell(w)$ can be obtained by extending maximal repeat $\ell(v)$ to the right \emph{and to the left}. Thus, for every arc $\gamma=(v,w)$ of the CDAWG, we store the first character of $\ell(\gamma)$ in a variable $\gamma.\mathtt{char}$, and we store the length of the right extension implied by $\gamma$ in a variable $\gamma.\mathtt{right}$. The length $\gamma.\mathtt{left}$ of the left extension implied by $\gamma$ can be computed by $w.\mathtt{length}-v.\mathtt{length}-\gamma.\mathtt{right}$. 
For every arc of the CDAWG that connects a maximal repeat $W$ to the sink, we store just $\gamma.\mathtt{char}$ and the starting position $\gamma.\mathtt{pos}$ of string $W \cdot \gamma.\mathtt{char}$ in $T$. The total space used by the CDAWG is $O(e_{T\#})$ words, and the number of runs in $\BWT_{T\#}$ can be shown to be $O(e_{T\#})$ as well \cite{belazzougui2015composite}. An alternative construction could use $\CDAWG_{\REV{T}\#}$ and $\RLBWT_{\REV{T}\#}$.

We use the RLBWT to count the number of occurrences of $P$ in $T$ in $O(m\log{\log{n}})$ time: if this number is not zero, we use the CDAWG to report all the $\mathtt{occ}$ occurrences of $P$ in $O(\mathtt{occ})$ time, using a technique already sketched in \cite{crochemore1997automata}. Specifically, since we know that $P$ occurs in $T$, we perform a blind search for $P$ in the CDAWG, as is typically done with PATRICIA trees. We keep a variable $i$, initialized to zero, that stores the length of the prefix of $P$ that we have matched so far, and we keep a variable $j$, initialized to one, that stores the starting position of $P$ inside the last maximal repeat encountered during the search. For every node $v$ in the CDAWG, we choose the arc $\gamma$ such that $\gamma.\mathtt{char}=P[i+1]$ in constant time using hashing, we increment $i$ by $\gamma.\mathtt{right}$, and we increment $j$ by $\gamma.\mathtt{left}$. If the search leads to the sink by an arc $\gamma$, we report $\gamma.\mathtt{pos}+j$ and we stop. If the search ends in a node $v$ that is associated with the maximal repeat $W$, we determine all the occurrences of $W$ in $T$ by performing a depth-first traversal of all nodes reachable from $v$ in the CDAWG , updating variables $i$ and $j$ as described above, and reporting $\gamma.\mathtt{pos}+j$ for every arc $\gamma$ that leads to the sink. The total number of nodes and arcs reachable from $v$ is $O(\mathtt{occ})$.

\section{Combining RLBWT and LZ factors in practice}

We implement\footnote{The source code of all our implementations is available at \cite{githubNicola1,githubNicola2}, and it is based on the SDSL library~\cite{gbmp2014sea}.} the combination of RLBWT and LZ factorization described in Section \ref{sec:compositeStringIndexes}, exploring a range of practical variants of decreasing size. Specifically, in addition to the version described in Section \ref{sec:compositeStringIndexes} (which we call \emph{full} in what follows), we implement a variant in which we drop $\RLBWT_{\REV{T}\#}$, simulating it with a bidirectional index (we call this variant \emph{bidirectional} in what follows), a variant in which we drop $\RLBWT_{\REV{T}\#}$, the four-sided range reporting data structure, and the subset of suffix tree nodes (we call this variant \emph{light} in what follows), and another variant in which we use a sparse version of the LZ parsing (we call this \emph{sparse} in what follows). Moreover, we design a number of practical optimization to speed up locate queries: see Appendix \ref{sec:heuristics}.

We implement all variants using a representation of the RLBWT that is more space-efficient than the one described in~\cite{SirenVMN08}. Recall that the latter is encoded as follows: they store one character per run in a string $H \in \Sigma^\runs$, they mark with a one the beginning of each run in a bitvector $V_{all}[0..n-1]$, and for every $c \in \Sigma$ they store the lengths of all runs of character $c$ consecutively in a bit-vector $V_c$: specifically, every $c$-run of length $k$ is represented in $V_c$ as $\mathtt{10}^{k-1}$. 
This representation allows one to map rank and access queries on $\BWT_{T\#}$ to rank, select and access queries on $H$, $V_{all}$, and $V_c$. By gap-encoding the bitvectors, this representation takes $\runs(2\log(n/\runs)+\log{\sigma})(1+o(1))$ bits of space. We reduce the multiplicative factor of term $\log(n/\runs)$ by storing in $V_{all}$ just one out of $1/\epsilon$ ones, where $0<\epsilon\leq 1$ is an arbitrary constant. It is easy to see that we are still able to answer all queries on the RLBWT by using the vectors $V_c$ to reconstruct the positions of the missing ones in $V_{all}$, using $\runs\big((1+\epsilon)\log(n/\runs)+\log{\sigma}\big)(1+o(1))$ bits of space, but query times are multiplied by a factor $1/\epsilon$. In all our experiments we set $\epsilon$ to 1/8. We represent $H$ as a Huffman-encoded string (\texttt{wt\_huff<>} in SDSL), and gap-encoded bitvectors with Elias-Fano (\texttt{sd\_vector<>} in SDSL).

\subsection{Full index}

The first variant is an engineered version of the data structure described in Section \ref{sec:compositeStringIndexes}. We store both $\RLBWT_{T\#}$ and $\RLBWT_{\REV{T}\#}$. A gap-encoded bitvector $\mathtt{end}[0..n-1]$ of $z\log(n/z)(1+o(1))$ bits marks the rank, among all the suffixes of $\REV{T}\#$, of every suffix $\REV{T}[i..n-1]\#$ such that $n-i-2$ is the last position of an LZ factor of $T$. Symmetrically, a gap-encoded bitvector $\mathtt{begin}[0..n-1]$ of $z\log(n/z)(1+o(1))$ bits marks the rank, among all the suffixes of $T\#$, of every suffix $T[i..n-1]\#$ such that $i$ is the first position of an LZ factor of $T$. 

Geometric range data structures (4-sided and 2-sided) are implemented as wavelet trees (\texttt{wt\_int<>} in SDSL). The 4-sided range data structure supports locating primary occurrences, by storing the permutation of the $z$ LZ factors of $T$, sorted lexicographically, in the order induced by the corresponding ones in $\mathtt{end}$: in other words, every character of the wavelet tree is the lexicographic rank of an LZ factor, among all the LZ factors of $T$. For locating primary occurrences, we need to label every point in the 4-sided range data structure with a text position. We allocate $\log{z}$ bits rather than $\log{n}$ bits to every such label, by using as label the rank of the corresponding one in array $\mathtt{begin}$, thus the data structure takes $2z\log{z}(1+o(1))$ bits of space. The 2-sided range data structure stores $z$ two-dimensional points whose coordinates are both in $[1..|T|]$. For locating secondary occurrences, every such point is again labeled with the rank of the corresponding one in array $\mathtt{begin}$. The wavelet tree that implements the 2-sided range data structure can only store points whose set of $x$ coordinates is $[1..z]$: 
we map text coordinates to this domain using a gap-encoded bitvector, which takes $z\log(n/z)(1+o(1))$ bits of space. Some coordinates could be repeated, since two LZ factors could share the same source start point: we keep track of duplicates using a succinct bitvector that takes $z+o(z)$ bits of space, in which a coordinate that is repeated $k>0$ times is encoded as $\mathtt{10}^{k-1}$. Overall, the 2-sided range data structure takes $z(2\log n + 1)(1+o(1))$ bits of space.

Finally, we need a way to compute the lexicographic range of a string among all the 
LZ factors of $T$. We implement a simpler and more space-efficient strategy than the one proposed in~\cite{belazzougui2015composite}. Specifically, recall that LZ factors are right-maximal substrings of $T\#$, or equivalently they are nodes of the suffix tree of $T\#$. Recall also that the BWT intervals of two nodes of the suffix tree of $T\#$ are either disjoint or contained in one another. We sort the BWT intervals of all LZ factors by the order induced by the pre-order traversal of the suffix tree of $T\#$: two distinct nonempty intervals $[i'..j']$ and $[i''..j'']$ are such that $[i'..j']<[i''..j'']$ iff $j'<i''$, or iff $[i''..j'']$ is contained in $[i'..j']$. The data structure is just the sorted array $V$ of such intervals, and it takes $2z\log{n}$ bits of space\footnote{The sequence of first positions of all intervals in the sorted array is non-decreasing, thus we use gap encoding to save $z\log{z}$ bits of space. This adds a multiplicative factor of $O(\log(n/z))$ to all query times. For clarity we describe just the simpler version in which intervals are encoded as $2z$ integers of $\log{n}$ bits each.}. Given the BWT interval $[i..j]$ of a string $W$, we find its lexicographic range among all sorted distinct LZ factors, in $O(\log{z})$ time, as follows: (1) we binary-search $V$ using the order described above, finding all intervals that are strictly smaller than $[i..j]$; (2) starting from the first interval in $V$ that is greater than or equal to $[i..j]$, we find all intervals in $V$ that equal $[i..j]$ or are contained in it, i.e. all intervals of factors that are either $W$ itself or a right-extension of $W$. This requires just one binary search, since all such intervals are contiguous in $V$.

In summary, the full index takes $\big( 6z\log{n} + 2(1+\epsilon)\runs\log(n/\runs) + 2\runs\log{\sigma} \big)\cdot(1+o(1))$ bits of space, and it supports count queries in $O(m\cdot (\log(n/\runs)+\log{\sigma}))$ time and locate queries in $O( (m + \occ)\cdot \log{n} )$ time.

\subsection{Bidirectional index}

We can drop $\RLBWT_{\REV{T}\#}$ and simulate it using just $\RLBWT_{T\#}$, by applying the synchronization step performed in bidirectional BWT indexes (see e.g. \cite{belazzougui2014linear} and references therein). This strategy penalizes the time complexity of locate queries, which becomes quadratic in the length of the pattern. Moreover, since in our implementation we store run-lengths separately for each character, a synchronization step requires $\sigma$ rank queries to find the number of characters smaller than a given character inside a BWT interval. This operation could be performed in $O(\log{\sigma})$ time if the string were represented as a wavelet tree. In summary, the bidirectional variant of the index takes $\big( 6z\log{n} + (1+\epsilon)\runs\log(n/\runs) + \runs\log{\sigma} \big)\cdot(1+o(1))$ bits of space, it supports count queries in $O(m\cdot (\log(n/\runs)+\log{\sigma}))$ time, and it supports locate queries in $O( m^2\sigma\log(n/\runs)+ (m + \occ)\cdot \log{n})$ time. 

\subsection{Light index with LZ sparsification}

Once we have computed the interval of the pattern in $\BWT_{T\#}$, we can locate all its primary occurrences by just forward-extracting at most $m$ characters for each occurrence inside the range: this is because every primary occurrence of the pattern overlaps with the last position of an LZ factor. We implement forward extraction with select queries on $\RLBWT_{T\#}$. This approach requires just $\RLBWT_{T\#}$, the 2-sided range data structure, a gap-encoded bitvector $\mathtt{end}_T$ that marks the last position of every LZ factor in the text, a gap-encoded bitvector $\mathtt{end}_{BWT}$ that marks the last position of every LZ factor in $\BWT_{T\#}$, and $z$ integers of $\log{z}$ bits each, connecting corresponding ones in $\mathtt{end}_{BWT}$ and in $\mathtt{end}_T$: this array plays the role of the sparse suffix array sampling used in RLCSA. 

We can reduce space even further by \emph{sparsifying the LZ factorization}. 
Intuitively, the factorization of a highly-repetitive collection of strings $T=T_1 T_2 \cdots T_k$, where $T_2,\dots,T_k$ are similar to $T_1$, is much denser 
inside $T_1$ than it is inside $T_2 \cdots T_k$, thus excluding long enough contiguous regions from the factorization (i.e. not outputting factors inside such regions) could reduce the number of factors in dense regions. Formally, let $d>0$, and consider the following generalization\footnote{The version of LZ77 considered in this paper is obtained by setting $d=0$, and by requiring the text to be virtually preceded by all its distinct characters.} of LZ77, denoted here by LZ77-$d$: we factor $T$ as $X_1Y_1X_2Y_2 \cdots X_{z_d}Y_{z_d}$, where $z_d$ is the size of the factorization, $Y_i \in \Sigma^d$ for all $i \in [1..z_d]$, and $X_i$ is the longest prefix of $X_iY_i \cdots X_{z_d}Y_{z_d}$ that appears at least twice in $X_1Y_1X_2Y_2 \cdots X_i$ To make the index described in this section work with LZ77-$d$, we need to sample the suffix array of $T\#$ at the lexicographic ranks that correspond to the last position of every $X_i$, and we need to redefine primary occurrences as those that are not fully contained inside an $X_i$. During locate we now need to extract $d$ additional characters before each occurrence of the pattern, in order to locate primary occurrences that start inside a $Y_i$. The 2-sided range data structure must also be built on the (sources of the) factors $X_1,\dots,X_{z_d}$. 
This implementation of the index takes $\big(z_d(3\log n + \log (n/z_d)) +(1+\epsilon)\runs\log(n/\runs) \big)\cdot(1+o(1))$ bits of space, it answers locate queries in $O((occ+1) \cdot (m+d) \cdot \log{n})$ time and count queries in $O(m(\log(n/\runs)+\log{\sigma}))$ time.

\section{Combining RLBWT and CDAWG in practice}

We implement\footnote{The source code of all our implementations is available at \cite{githubMathieu}.} the combination of RLBWT and CDAWG described in Section \ref{sec:compositeStringIndexes}, and we study the effect of two representations of the CDAWG in memory. In the first representation, the graph is encoded as a sequence of variable-length integers: every integer is represented as a sequence of bytes, in which the seven least significant bits of every byte are used to encode the integer, and the most significant bit flags the last byte of the integer. Nodes are stored in the sequence according to their topological order in the graph obtained from the CDAWG by inverting the direction of all arcs: to encode a pointer from a node $v$ to its successor $w$ in the CDAWG, we store the difference between the first byte of $v$ and the first byte of $w$ in the sequence. If $w$ is the sink, such difference is replaced by a shorter code. We choose to store the length of the maximal repeat that corresponds to each node, rather than the offset of $\ell(v)$ inside $\ell(w)$ for every arc $(v,w)$, since such lengths are short and smaller than the number of arcs in practice.

In the second encoding we exploit the fact that the subgraph of the suffix tree of $T\#$ induced by maximal repeats is a spanning tree of $\CDAWG_{T\#}$ (see Section \ref{sec:strings}). Specifically, we encode such spanning tree with the balanced parenthesis scheme described in \cite{Munro01}, and we resolve the arcs of the CDAWG that belong to the tree using corresponding tree operations. Such operations work on node identifiers, thus we need to convert node identifiers to the first byte in the byte sequence of the CDAWG, and vice versa. For this, we encode the monotone sequence of the first byte of all $n$ nodes in the byte sequence using the quasi-succinct representation by Elias and Fano, which uses at most $2+\log(N/n)$ bits per starting position, where $N$ is the number of bytes in the byte sequence \cite{elias1975complexity}.

Finally, we observe that classical CDAWG construction algorithms (e.g. the online algorithm described in \cite{CrochemoreV97}) are not space-efficient, and we design linear-time algorithms that build a representation of the CDAWG from $\BWT_{T\#}$ or $\BWT_{\REV{T}\#}$, using optimal additional space. Specifically, let $\mathtt{enumerateLeft}(i,j)$ be a function that returns the set of distinct characters that appear in $\BWT_{T\#}[i..j]$, not necessarily in lexicographic order. We prove the following lemmas in Appendix \ref{appendix:cdawg1} and \ref{appendix:cdawg2}:

\begin{lemma} \label{lemma:cdawg1}
Let $T \in [1..\sigma]^{n-1}\#$, where $\#=0$. Given a representation of $\BWT_{T\#}$ that answers $\mathtt{enumerateLeft}$ in time $t_{EL}$ per element in its output, and LF in time $t_{LF}$, we can build the topology of $\CDAWG_{T\#}$, as well as the first character and the length of the label of each arc, in randomized $O(n(t_{EL}+t_{LF}))$ time and zero space in addition to the input and the output. 
\end{lemma}

\begin{lemma} \label{lemma:cdawg2}
Let $T \in [1..\sigma]^{n-1}\#$, where $\#=0$. Given a representation of $\BWT_{\REV{T}\#}$ that answers $\mathtt{enumerateLeft}$ in time $t_{EL}$ per element in its output, and LF in time $t_{LF}$, we can build the topology of $\CDAWG_{T\#}$, as well as the first character and the length of the label of each arc, in randomized $O(n(t_{EL}+t_{LF}))$ time and $O(\sigma^{2}\log^{2}{n})$ bits of space in addition to the input and the output.
\end{lemma}

\section{Experimental results} \label{sec:experiments}

We test our implementations on five DNA datasets from the Pizza\&Chili repetitive corpus \cite{pizzachili}, which include the whole genomes of approximately 36 strains of the same eukaryotic species (``Saccharomyces cerevisiae'' and ``Saccharomyces paradoxus'' in the plots), a collection of 23 and approximately 78 thousand substrings of the genome of the same bacterium (respectively ``Escherichia coli'' and ``Haemophilus influenzae''), and an artificially repetitive string obtained by concatenating a hundred mutated copies of the same substring of the human genome (denoted by ``pseudo-real'' in the plots)\footnote{Compressing such files with $\mathtt{p7zip}$ \cite{p7zip}, an implementation of LZ77 with large window, makes the uncompressed files between 21 and 370 times bigger than the corresponding compressed files \cite{pizzachili}.}. 
We compare our results to the FM index implementation in SDSL \cite{gbmp2014sea} with sampling rate $2^i$ for $i \in [5..10]$ (represented by black circles in all plots), to an implementation of RLCSA\footnote{We compile the sequential version with \texttt{PSI\_FLAGS} and \texttt{SA\_FLAGS} turned off (thus, a gap-encoded bitvector rather than a succinct bitvector is used to mark sampled positions in the suffix array). The block size of psi vectors (\texttt{RLCSA\_BLOCK\_SIZE}) is 32 bytes.} \cite{adamnovak} with the same sampling rates (triangles in the plots), to the five variants in the implementation of the LZ77 index described in \cite{kreft2010self} (squares), and to a recent implementation of the compressed hybrid index \cite{valenzuela2016chico} (diamonds). The FM index uses RRR bitvectors in its wavelet tree. For brevity, we call LZ1 the implementation of the LZ77 index that uses suffix trie and reverse trie. For each process, and for each pattern length $2^i$ for $i \in [3..10]$, we measure the maximum resident set size and the number of CPU seconds that the process spends in user mode\footnote{We perform all experiments on a single core of a 6-core, 2.50 GHz, Intel Xeon E5-2640 processor, with access to 128GiB of RAM and running CentOS 6.3. We measure resources with GNU Time 1.7, and we compile with GCC 5.3.0.}, both for locate and for count queries, discarding the time for loading the indexes and averaging our measurements over one thousand patterns\footnote{We generate random patterns that contain just characters in $\{\mathtt{a},\mathtt{c},\mathtt{g},\mathtt{t}\}$ using the \texttt{genpatterns} tool from the Pizza\&Chili corpus \cite{pizzachili}.}.

\begin{figure}
\begin{center}
\includegraphics[width=1\textwidth]{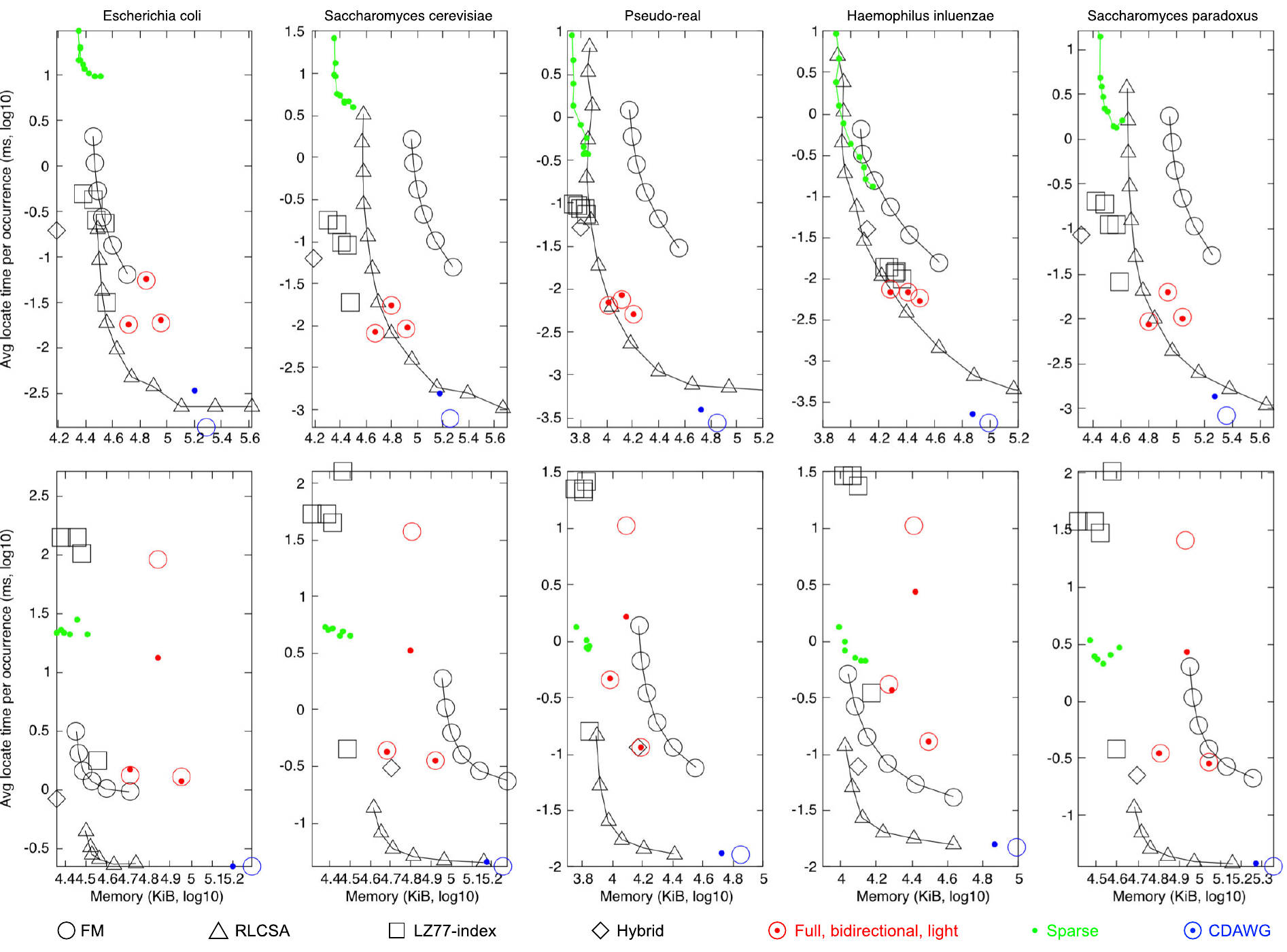}
\caption{Space-time tradeoffs of our indexes (color) and of the state of the art (black). Top row: patterns of length 16. Bottom row: patterns of length 512. For more clarity, RLCSA and sparse index are tested also on additional configurations not mentioned in Section \ref{sec:experiments}.
\label{fig:tradeoffs}
}
\end{center}
\end{figure}

\begin{figure}
\begin{center}
\includegraphics[width=1\textwidth]{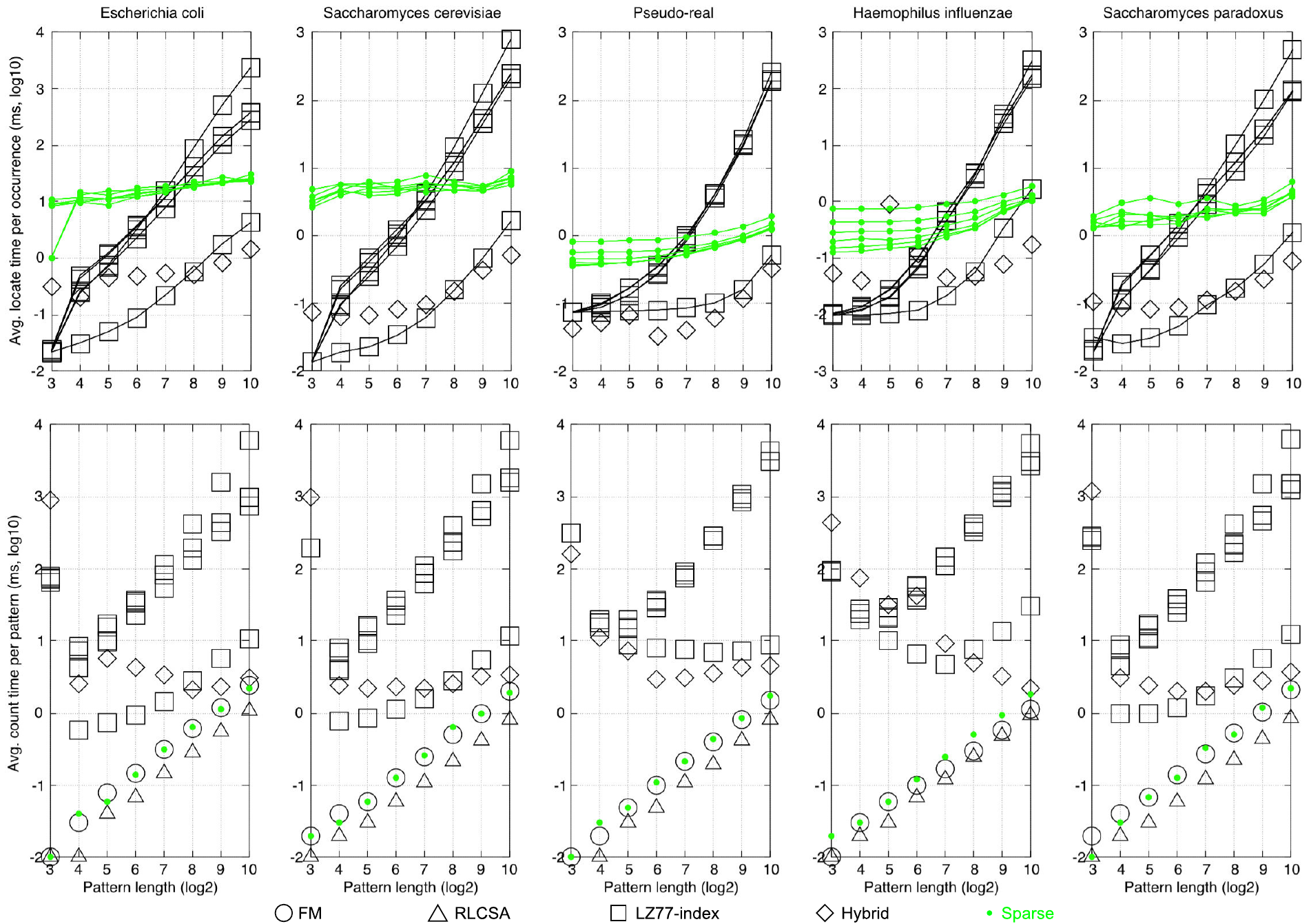}
\caption{Locate time per occurrence (top) and count time per pattern (bottom), as a function of pattern length, for the sparse index with skip rate $2^i$, $i \in [5..10]$, the LZ77 index, and the hybrid index. Count plots show also the FM index and RLCSA.
\label{fig:countLocate}
}
\end{center}
\end{figure}

We observe two distinct regimes for locate queries, corresponding to short patterns (shorter than approximately 64) and to long patterns, respectively (Figures \ref{fig:tradeoffs} and \ref{fig:countLocate}). Our full, bidirectional and light index implementations (red circles in all plots) do not achieve any new useful tradeoff, in any dataset, neither with short nor with long patterns. 
As expected, the running time per pattern of the bidirectional index depends quadratically on pattern length, but we observe a superlinear growth for the light index as well. 
The optimizations described in Appendix \ref{sec:heuristics} (red dots) are effective only for the bidirectional index, their effectiveness increases with pattern length, and they manage to shave up to 80\% of running time with patterns of length 1024. 
The size of the bidirectional index on disk is on average 20\% 
smaller than the size of the full index on disk, and the size of the light index on disk is approximately 20\% 
smaller than the size of the bidirectional index on disk.

\begin{figure}
\begin{center}
\includegraphics[width=1\textwidth]{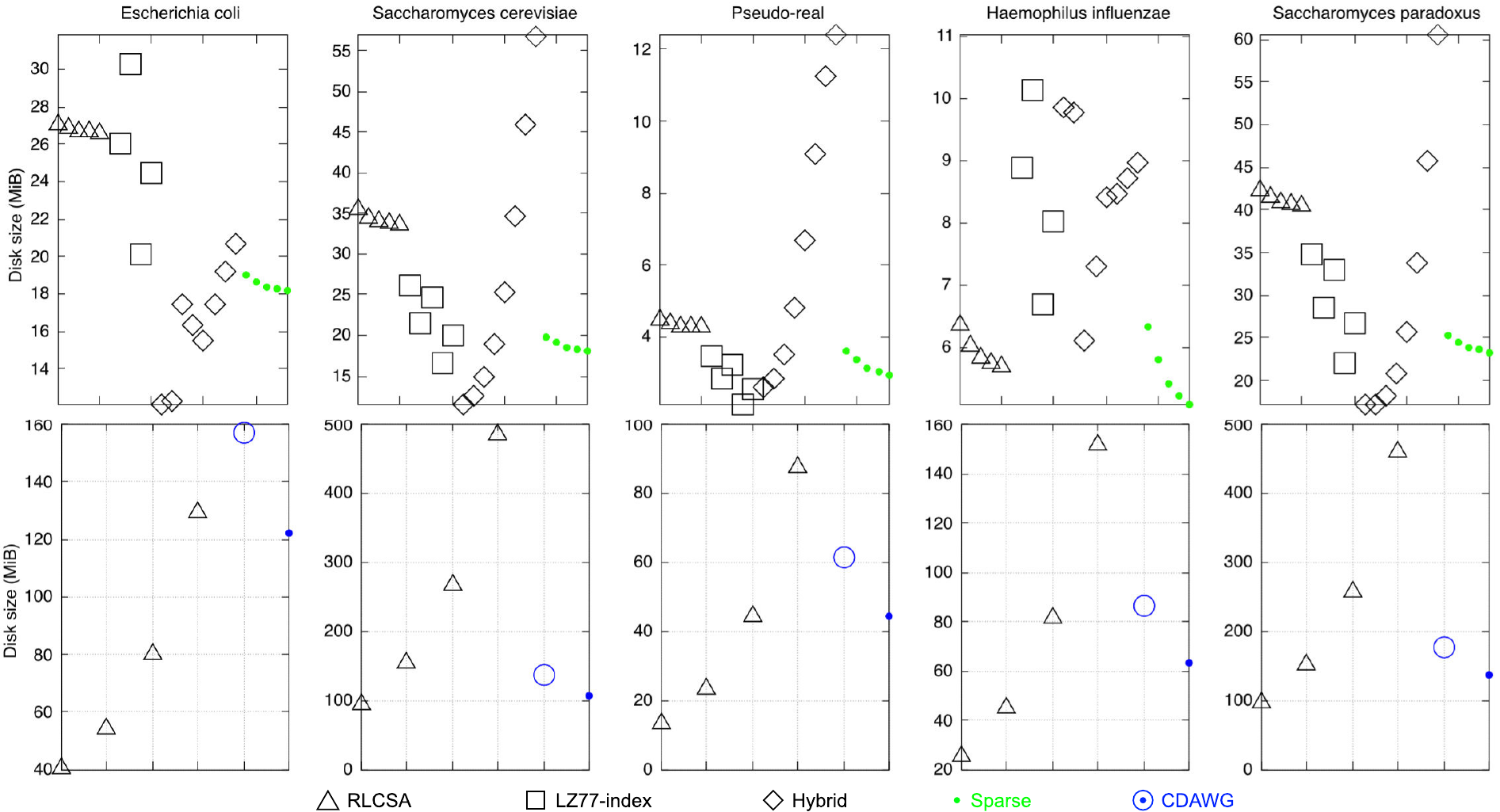}
\caption{(Top) Disk size of the sparse index with skip rate $2^i$, $i \in [10..15]$, compared to the hybrid index with maximum pattern length $2^i$, $i \in [3..10]$, the LZ77 index, and RLCSA with sampling rate $2^i$, $i \in [10..15]$. (Bottom) Disk size of the CDAWG compared to RLCSA with sampling rate $2^i$, $i \in [2..5]$.
\label{fig:disk}
}
\end{center}
\end{figure}

We experiment with skipping $2^i$ characters before opening a new phrase in the light index with LZ sparsification (green in all plots), where $i \in [5..10]$. The size of the sparse index with skip rate $2^5$ on disk is approximately 35\% smaller than the size of the light index on disk. With short patterns, the memory used by the sparse index becomes smaller than RLCSA and comparable to the LZ index, but its running time per occurrence is between one and two orders of magnitude greater than the LZ index and comparable to RLCSA with sampling rates equal to or greater than 2048 (Figure \ref{fig:tradeoffs}, top). \emph{With long patterns, however, the sparse index becomes between one and two orders of magnitude faster than all variants of the LZ index, except variant LZ1, while using comparable memory}. As a function of pattern length, the running time per occurrence of the sparse index grows more slowly than the running time of LZ1, suggesting that \emph{the sparse index becomes as fast as LZ1 for patterns of length between 1024 and 2048} (Figure \ref{fig:countLocate}, top). The sparse index is approximately 1.5 orders of magnitude slower than the hybrid index, but since the size of the hybrid index depends on maximum pattern length, \emph{the sparse index becomes smaller than the hybrid index for patterns of length between 64 and 128}, and possibly even shorter (Figure \ref{fig:disk}, top). As expected, the sparse index is faster than both the LZ index and the hybrid index in count queries, especially for short patterns: specifically, \emph{the sparse index is between two and four orders of magnitude faster than all variants of the LZ index}, with the largest difference for patterns of length 8 (Figure \ref{fig:countLocate}, bottom). The difference between the sparse index and variant LZ1 shrinks as pattern length increases. Similar trends hold for the hybrid index. The full, bidirectional and light indexes show similar count times as the sparse index.

\begin{figure}
\begin{center}
\includegraphics[width=1\textwidth]{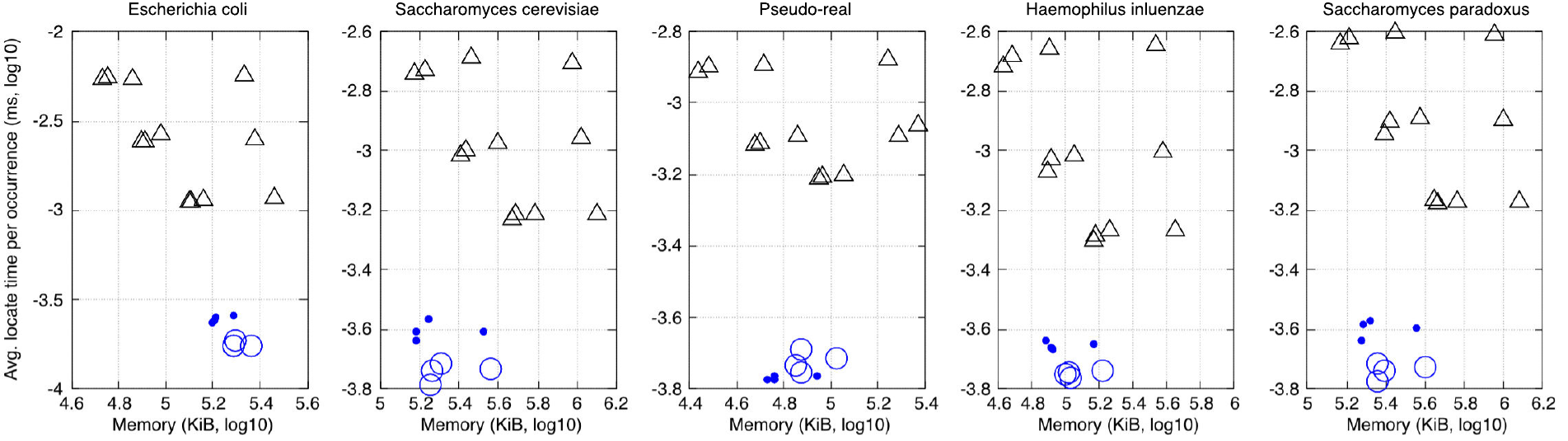}
\caption{Space-time traeoffs of the CDAWG (blue) compared to RLCSA (triangles) with sampling rate $2^i$, $i \in [3..5]$. Patterns of length 8, 6, 4, 2 (from left to right).
\label{fig:cdawg}
}
\end{center}
\end{figure}


The disk size of the CDAWG is comparable to the disk size of RLCSA with sampling rate between 4 and 8 (Figure \ref{fig:disk}, bottom). Using the succinct representation of the CDAWG (blue dots in all plots) shaves between 20\% and 30\% of the disk size and resident set of the non-succinct representation (blue circles). However, using the non-succinct representation shaves between 20\% and 80\% of the time of the succinct representation, depending on dataset and pattern length. Using the CDAWG to answer locate queries does not achieve any new tradeoff with long patterns (Figure \ref{fig:tradeoffs}, bottom). However, \emph{with short patterns the running time per occurrence of the CDAWG is between 4 and 10 times smaller than the running time per occurrence of a version of RLCSA that uses comparable memory, and with patterns of length two the CDAWG achieves speedups even greater than 10}.


\subparagraph*{Acknowledgements.}

We thank Miguel \'{A}ngel Mart\'{i}nez for providing implementations of the variants described in \cite{kreft2010self}, and Daniel Valenzuela for providing the implementation described in \cite{valenzuela2016chico}.

\section{Future work}

Designing indexes for repetitive texts is an increasingly active field and it is beyond the scope of this paper to review all the recent proposals --- see, e.g.,~\cite{FGHP14,GGKNP14,TTS14,valenzuela2016chico} --- especially since some have not been implemented. We would like to draw attention to an index described simultaneously and independently by Do et al.~\cite{DJSS14} and Gagie et al.~\cite{GGKNP12}, however, because we think it can be improved and made competitive in practice.

Their index is intended for collections of many similar strings, such as databases of genomes from the same species.  The main idea is to choose one of the strings as a reference and build an FM-index for the Relative Lempel-Ziv parse~\cite{KPZ10} of the entire dataset with respect to that reference, treating the phrases as meta-characters.  Using FM-indexes for the reference and its reverse and some auxiliary data structures, we can apply dynamic programming to quickly compute all the ways any given pattern can be decomposed into the suffix of a phrase (possibly empty), a sequence of complete phrases, and the prefix of a phrase (possibly empty).  This is possible because the dictionary of potential phrases --- i.e., all the substrings of the reference --- is fixed, albeit very large, and does not change while we parse (in contrast to the dictionaries for LZ77 and LZ78).

Using the FM-index for the parse and some more auxiliary data structures, we can quickly find whether and where any of the possible decompositions of the pattern occur in the the parse of the dataset.  These occurrences correspond to occurrences of the pattern that cross phrase boundaries.  We can then quickly find all the other occurrences of the pattern in the dataset.  Unfortunately, if there are many distinct phrases in the parse, the FM-index for it may not compress well.  This raises the question of how we can reduce the number of distinct phrases without increasing the number of phrases too much.

Suppose we build a compressed $k$th-order de Bruijn graph for the collection of strings (i.e., a $k$th-order de Bruijn graph in which we have collapsed every maximal path whose internal nodes have in- and out-degree 1); assign each of its edges with a distinct meta-character; consider each string as a walk on the graph and, for each edge the walk crosses, replacing the substring causing us to cross that edge by the edge's meta-character.  This results in a parse in which the number of distinct phrases is the number of edges in the graph (assuming each of the strings in the collection has length at least $k$).  Notice also that, for any pattern of length at least $k$, the pattern corresponds to at most one walk on the uncompressed graph (which may start and finish in the middle of edges in the compressed graph); if there is no such walk, then the pattern does not occur in the collection.  Using a de Bruijn graph also removes the need for choosing a reference and using range reporting.

We may be able to improve the compression by, first, removing low-frequency edges in the original de Bruijn graph before compressing it and, second, replacing substrings only when they cause us to cross high-frequency edges with sufficiently long edge labels in the the compressed de Bruijn graph.  With this modification, a pattern may appear in the collection even if it does not correspond to a walk on the uncompressed graph.  However, if a substring causes us to cross a high-frequency edge with a long edge label in the compressed graph, then we can certainly replace that substring by the edge's meta-character.  It follows that, for any pattern of length at least $k$, we need perform at most four searches in the FM-index for the parse to determine whether and where that pattern occurs in the collection.

We plan to implement and test this modification of Do et al.'s and Gagie et al.'s index soon and report the results in a future paper.

\subparagraph*{Acknowledgements.}

We thank Miguel \'{A}ngel Mart\'{i}nez for providing implementations of the variants described in \cite{kreft2010self}, and Daniel Valenzuela for providing the implementation described in \cite{valenzuela2016chico}.




\begin{thebibliography}{10}

\bibitem{arroyuelo2012stronger}
Diego Arroyuelo, Gonzalo Navarro, and Kunihiko Sadakane.
\newblock Stronger {Lempel-Ziv} based compressed text indexing.
\newblock {\em Algorithmica}, 62(1-2):54--101, 2012.

\bibitem{belazzougui2014linear}
Djamal Belazzougui.
\newblock Linear time construction of compressed text indices in compact space.
\newblock In {\em Proceedings of the 46th Annual ACM Symposium on Theory of
  Computing}, pages 148--193. ACM, 2014.

\bibitem{belazzougui2015composite}
Djamal Belazzougui, Fabio Cunial, Travis Gagie, Nicola Prezza, and Mathieu
  Raffinot.
\newblock Composite repetition-aware data structures.
\newblock In {\em Combinatorial Pattern Matching}, pages 26--39. Springer,
  2015.

\bibitem{blumer1987complete}
Anselm Blumer, Janet Blumer, David Haussler, Ross McConnell, and Andrzej
  Ehrenfeucht.
\newblock Complete inverted files for efficient text retrieval and analysis.
\newblock {\em Journal of the ACM}, 34(3):578--595, 1987.

\bibitem{chan2011orthogonal}
Timothy~M Chan, Kasper~Green Larsen, and Mihai P{\u{a}}tra{\c{s}}cu.
\newblock Orthogonal range searching on the {RAM}, revisited.
\newblock In {\em Proceedings of the twenty-seventh annual symposium on
  computational geometry}, pages 1--10. ACM, 2011.

\bibitem{crochemore1997automata}
Maxime Crochemore and Christophe Hancart.
\newblock Automata for matching patterns.
\newblock In {\em Handbook of formal languages}, pages 399--462. Springer,
  1997.

\bibitem{CrochemoreV97}
Maxime Crochemore and Renaud V\'erin.
\newblock Direct construction of compact directed acyclic word graphs.
\newblock In Alberto Apostolico and Jotun Hein, editors, {\em CPM}, volume 1264
  of {\em Lecture Notes in Computer Science}, pages 116--129. Springer, 1997.

\bibitem{DJSS14}
Huy~Hoang Do, Jesper Jansson, Kunihiko Sadakane, and Wing{-}Kin Sung.
\newblock Fast relative lempel-ziv self-index for similar sequences.
\newblock {\em Theoretical Computer Science}, 532:14--30, 2014.

\bibitem{elias1975complexity}
Peter Elias and Richard~A Flower.
\newblock The complexity of some simple retrieval problems.
\newblock {\em Journal of the ACM (JACM)}, 22(3):367--379, 1975.

\bibitem{FGHP14}
Hector Ferrada, Travis Gagie, Tommi Hirvola, and Simon J.~Puglisi.
\newblock Hybrid indexes for repetitive datasets.
\newblock Philosophical Transactions of the Royal Society of London A, 372(2016), 2014.

\bibitem{pizzachili}
Paolo Ferragina and Gonzalo Navarro.
\newblock {Pizza\&Chili repetitive corpus}.
\newblock \url{http://pizzachili.dcc.uchile.cl/repcorpus.html}.
\newblock Accessed: 2016-04-10.

\bibitem{GGKNP12}
Travis Gagie, Pawel Gawrychowski, Juha K{\"{a}}rkk{\"{a}}inen, Yakov Nekrich,
  and Simon~J. Puglisi.
\newblock A faster grammar-based self-index.
\newblock In {\em Proceedings of the 6th Conference on Language and Automata
  Theory and Applications}, pages 240--251, 2012.

\bibitem{gagie2014lz77}
Travis Gagie, Pawe{\l} Gawrychowski, Juha K{\"a}rkk{\"a}inen, Yakov Nekrich,
  and Simon~J Puglisi.
\newblock {LZ77}-based self-indexing with faster pattern matching.
\newblock In {\em {LATIN} 2014: Theoretical Informatics}, pages 731--742.
  Springer, 2014.

\bibitem{GGKNP14}
Travis Gagie, Pawel Gawrychowski, Juha K{\"{a}}rkk{\"{a}}inen, Yakov Nekrich,
  and Simon~J. Puglisi.
\newblock Lz77-based self-indexing with faster pattern matching.
\newblock In {\em Proceedings of the 11th Latin American Symposium on
  Theoretical Informatics}, pages 731--742, 2014.

\bibitem{gbmp2014sea}
Simon Gog, Timo Beller, Alistair Moffat, and Matthias Petri.
\newblock From theory to practice: Plug and play with succinct data structures.
\newblock In {\em 13th International Symposium on Experimental Algorithms, (SEA
  2014)}, pages 326--337, 2014.

\bibitem{karkkainen1996lempel}
Juha K{\"a}rkk{\"a}inen and Esko Ukkonen.
\newblock {Lempel-Ziv} parsing and sublinear-size index structures for string
  matching.
\newblock In {\em Proc. 3rd South American Workshop on String Processing
  (WSP'96}, pages 141--155, 1996.

\bibitem{kreft2010self}
Sebastian Kreft.
\newblock Self-index based on lz77.
\newblock Master's thesis, Department of Computer Science, University of Chile,
  2010.

\bibitem{kreft2013compressing}
Sebastian Kreft and Gonzalo Navarro.
\newblock On compressing and indexing repetitive sequences.
\newblock {\em Theoretical Computer Science}, 483:115--133, 2013.

\bibitem{KPZ10}
Shanika Kuruppu, Simon~J. Puglisi, and Justin Zobel.
\newblock Relative lempel-ziv compression of genomes for large-scale storage
  and retrieval.
\newblock In {\em Proceedings of the 17th Symposium on String Processing and
  Information Retrieval}, pages 201--206, 2010.

\bibitem{makinen2005succinct1}
Veli M{\"a}kinen and Gonzalo Navarro.
\newblock Succinct suffix arrays based on run-length encoding.
\newblock In {\em Combinatorial Pattern Matching}, pages 45--56. Springer,
  2005.

\bibitem{MakinenNSV10}
Veli M{\"{a}}kinen, Gonzalo Navarro, Jouni Sir{\'{e}}n, and Niko
  V{\"{a}}lim{\"{a}}ki.
\newblock Storage and retrieval of highly repetitive sequence collections.
\newblock {\em Journal of Computational Biology}, 17(3):281--308, 2010.

\bibitem{morrison1968patricia}
Donald~R Morrison.
\newblock Patricia—practical algorithm to retrieve information coded in
  alphanumeric.
\newblock {\em Journal of the ACM (JACM)}, 15(4):514--534, 1968.

\bibitem{Munro01}
J.~Ian Munro and Venkatesh Raman.
\newblock Succinct representation of balanced parentheses and static trees.
\newblock {\em SIAM J. Comput.}, 31(3):762--776, March 2002.
\newblock URL: \url{http://dx.doi.org/10.1137/S0097539799364092}, \href
  {http://dx.doi.org/10.1137/S0097539799364092}
  {\path{doi:10.1137/S0097539799364092}}.

\bibitem{adamnovak}
Adam Novak.
\newblock Convenient repository for/fork of the {RLCSA} library.
\newblock \url{https://github.com/adamnovak/rlcsa}.
\newblock Accessed: 2016-04-10.

\bibitem{p7zip}
Igor Pavlov.
\newblock {P7ZIP} home.
\newblock \url{http://p7zip.sourceforge.net}.
\newblock Accessed: 2016-04-10.

\bibitem{githubNicola1}
Nicola Prezza.
\newblock \texttt{lz-rlbwt}: {R}un-length compressed {B}urrows-{W}heeler
  transform with {LZ77} suffix array sampling.
\newblock \url{https://github.com/nicolaprezza/lz-rlbwt}.
\newblock Accessed: 2016-04-10.

\bibitem{githubNicola2}
Nicola Prezza.
\newblock \texttt{lz-rlbwt-sparse}: {R}un-length compressed {B}urrows-{W}heeler
  transform with sparse {LZ77} suffix array sampling.
\newblock \url{https://github.com/nicolaprezza/lz-rlbwt-sparse}.
\newblock Accessed: 2016-04-10.

\bibitem{githubMathieu}
Mathieu Raffinot.
\newblock \texttt{locate-cdawg}: {R}eplacing sampling by {CDAWG} localisation
  in {BWT} indexing approaches.
\newblock \url{https://github.com/mathieuraffinot/locate-cdawg}.
\newblock Accessed: 2016-04-10.

\bibitem{Raffinot2001}
Mathieu Raffinot.
\newblock On maximal repeats in strings.
\newblock {\em Information Processing Letters}, 80(3):165--169, 2001.

\bibitem{SirenVMN08}
Jouni Sir{\'{e}}n, Niko V{\"{a}}lim{\"{a}}ki, Veli M{\"{a}}kinen, and Gonzalo
  Navarro.
\newblock Run-length compressed indexes are superior for highly repetitive
  sequence collections.
\newblock In {\em String Processing and Information Retrieval, 15th
  International Symposium, {SPIRE} 2008, Melbourne, Australia, November 10-12,
  2008.}, pages 164--175, 2008.

\bibitem{TTS14}
Yoshimasa Takabatake, Yasuo Tabei, and Hiroshi Sakamoto.
\newblock Improved esp-index: {A} practical self-index for highly repetitive
  texts.
\newblock In {\em Proceedings of the 13th Symposium on Experimental
  Algorithms}, pages 338--350, 2014.

\bibitem{valenzuela2016chico}
Daniel Valenzuela.
\newblock {CHICO}: A compressed hybrid index for repetitive collections.
\newblock In {\em Proceedings of the fifteenth International Symposium on
  Experimental Algorithms (SEA 2016)}, Lecture Notes in Computer Science.
  Springer, June 2016.

\bibitem{Wi83}
Dan~E Willard.
\newblock Log-logarithmic worst-case range queries are possible in space
  {Theta}(n).
\newblock {\em Information Processing Letters}, 17(2):81--84, 1983.

\bibitem{ziv1977universal}
Jacob Ziv and Abraham Lempel.
\newblock A universal algorithm for sequential data compression.
\newblock {\em IEEE Transactions on information theory}, 23(3):337--343, 1977.

\end{thebibliography}


\appendix

\section{Proof of Lemma \ref{lemma:cdawg1}} \label{appendix:cdawg1}

We use the algorithm described in \cite{belazzougui2014linear} to enumerate a representation of every node of $\ST_{T\#}$ by performing a depth-first traversal of the \emph{suffix link tree} of $T\#$. Such algorithm works in $O(n \cdot t_{EL})$ time and in $O(\sigma^{2}\log^{2}n)$ bits of working space\footnote{The enumeration algorithm described in \cite{belazzougui2014linear} uses the stack trick to fit the working space in $O(\sigma^{2}\log^{2}n)$ bits. Without such trick, the working space would be proportional to the largest number of left-extensions of maximal repeats that lie in the same path of the suffix-link tree of $T\#$, which can be charged to the output of Lemma \ref{lemma:cdawg1}.}
, and it provides, for each node $v$, its interval in $\BWT_{T\#}$ and the length of its label, as well as the list of all its children, and for every such child $w$, the interval of $w$ in $\BWT_{T\#}$ and the first character of the label of edge $(v,w)$ in $\ST_{T\#}$. 

Since the label of every node of the CDAWG is a maximal repeat of $T\#$, the set of nodes of the CDAWG (excluding the sink $\Omega$) is in one-to-one correspondence with a subset of the nodes of the suffix tree. Specifically, a node $v$ of $\ST_{T\#}$ corresponds to a node $v'$ of the CDAWG if and only if $\ell(v)$ is a left-maximal substring of $T\#$. We can check the left-maximality of $\ell(v)$ by counting the number of distinct characters in the BWT interval of $v$. Every time this number is greater than one, we have discovered a new node $v'$ of the CDAWG, and we assign to it a unique identifier by incrementing a global counter. Then, we scan every child $w$ of $v$ in $\ST_{T\#}$, and we store in a hash table a tuple $(i,j,v'.\mathtt{id},|\ell(v)|,c)$, where $[i..j]$ is the interval of $w$ and is used as key, and $v'.\mathtt{id}$ is the unique number assigned to $v'$. Note that every quadruplet we insert in the hash table has a unique key. If the BWT interval of $w$ is of length one, then $w$ is a leaf, an arc connects $v'$ in the CDAWG to the sink $\Omega$ with character $c$, and $c$ is the first character of the label of edge $(v,w)$ in $\ST_{T\#}$. We compute the starting position of $\ell(v) \cdot c$ for of all such arcs $(v',\Omega)$ in the CDAWG in batch, by inverting $\BWT_{T\#}$ and querying the hash table. The hash table can be implemented to support both insertion and querying in $O(1)$ randomized time. BWT inversion takes $O(n \cdot t_{LF})$ time and an amount of memory that can be charged to the output.

To build all arcs of the CDAWG that are not directed to the sink, we perform another traversal of the suffix-link tree, in the same order as the first traversal. Assume that, during this second traversal, we enumerate a node $w$ whose BWT interval is present in the hash table: then, $w$ is the child of a node $v$ of the suffix tree that corresponds to a node $v'$ of the CDAWG. If $w$ corresponds to a node $w'$ of the CDAWG as well, i.e. if $\ell(w)$ if a left-maximal substring of $T\#$, we add arc $(v',w')$ to the CDAWG. Otherwise $w$ is not left-maximal, thus the CDAWG must contain arc $(v',u')$ where $\ell(u') = W \cdot \ell(w)$ is the shortest left-extension of $\ell(w)$ to be left-maximal, and the corresponding node $u$ in the suffix tree can be reached from $w$ by a unary path in the suffix-link tree. Thus, we keep an auxiliary buffer, initially empty. Every time we encounter a node $w$ of the suffix tree whose interval $[i..j]$ is such that a tuple $(i,j,v'.\mathtt{id},|\ell(v')|,c)$ exists in the hash table, we append to the buffer tuple $(v'.\mathtt{id},|\ell(w)|-|\ell(v')|,c)$. Moreover, if $w$ is left-maximal, we empty the buffer and we transform every tuple $(v'.\mathtt{id},k,c)$ in the buffer into an arc $(v'.\mathtt{id},w'.\mathtt{id},k,c)$ of the CDAWG. The size of such buffer can be charged to the output.

\section{Proof of Lemma \ref{lemma:cdawg2}} \label{appendix:cdawg2}

We proceed as in Appendix \ref{appendix:cdawg1}, traversing the suffix-link tree of $\REV{T}\#$, and enumerating the intervals in $\BWT_{\REV{T}\#}$ of every node of the suffix tree of $\REV{T}\#$, i.e. of every right-maximal substring of $\REV{T}\#$. Once we detect that a node $v$ is also left-maximal in $\REV{T}\#$, we create a new node $v'$ of the CDAWG, we assign a new unique identifier to it, we enumerate all the \emph{left-extensions} $c \cdot \ell(v)$ in $\REV{T}\#$, and we push tuple $(i,j,v'.\mathtt{id},|\ell(v)|,c)$ in a hash table, where $[i..j]$ is the interval of $c \cdot \ell(v)$ in $\BWT_{\REV{T}\#}$ and is used as key, and $v'.\mathtt{id}$ is the unique number assigned to $v'$. If the BWT interval of $c \cdot \ell(v)$ is of length one, an arc connects $v'$ to the sink $\Omega$ with character $c$ in the CDAWG. We compute the starting position in $\REV{T}\#$ of $c \cdot \ell(v)$ for of all such arcs $(v',\Omega)$ in the CDAWG in batch, by inverting $\BWT_{\REV{T}\#}$ and querying the hash table. Note that, since we are possibly pushing the intervals in $\BWT_{\REV{T}\#}$ of the destinations of \emph{implicit Weiner links} in the suffix tree of $\REV{T}\#$, the hash table has to allow the presence of distinct tuples with the same key.

To build all arcs of the CDAWG that are not directed to the sink, we perform another traversal of the suffix-link tree of $\REV{T}\#$, in the same order as the first traversal. Assume that, during this second traversal, we enumerate a node $w$ of the suffix tree of $\REV{T}\#$ whose interval in $\BWT_{\REV{T}\#}$ is present in the hash table in a set of tuples $\mathcal{T}=\{(i,j,v'.\mathtt{id},|\ell(v)|,c)$\}. If $w$ corresponds to a node $w'$ of the CDAWG as well, i.e. if $\ell(w)$ if a left-maximal substring of $\REV{T}\#$, we add arc $(v',w')$ to the CDAWG for every tuple in $\mathcal{T}$. Otherwise $w$ is not left-maximal in $\REV{T}\#$, thus the CDAWG must contain arc $(v',u')$ for every tuple in $\mathcal{T}$, where $\ell(u') = W \cdot \ell(w)$ is the shortest left-extension of $\ell(w)$ to be left-maximal in $\REV{T}\#$, and the corresponding node $u$ in the suffix tree of $\REV{T}\#$ can be reached from $w$ by a unary path in the suffix-link tree of $\REV{T}\#$. Thus, we keep an auxiliary buffer, initially empty. Every time we encounter a node $w$ of the suffix tree of $\REV{T}\#$ whose interval $[i..j]$ is such that a set of tuples $\mathcal{T}$ exists in the hash table, we append to the buffer a corresponding set of tuples $\{(v'.\mathtt{id},|\ell(w)|,c)\}$. Moreover, if $w$ is left-maximal in $\REV{T}\#$, we empty the buffer and we transform every tuple $(v'.\mathtt{id},k,c)$ in the buffer into an arc $(v'.\mathtt{id},w'.\mathtt{id},|\ell(w)|-k+1,c)$ of the CDAWG. The size of such buffer can be charged to the output.

\section{Speeding up locate queries on indexes based on RLBWT and LZ factors} \label{sec:heuristics}

On indexes based on RLBWT and LZ factors, locate queries can be further engineered in a number of ways:
\begin{enumerate}
\item Thanks to the RLBWT, we know the total number of occurrences of $P$ in $T$ before starting to locate them: thus, we can stop locating as soon as we have found all occurrences.

\item We could add a compressed bitvector $\mathtt{first}[1..n]$ that flags a position $i$ iff $\SA_{T\#}[i]=p_j$ for some $j \in [1..z]$. As we backward-search $P$ in $\BWT_{T\#}$, we could mark the positions $k \in [2..m]$ such that the interval of $P[k..m]$ in $\BWT_{T\#}$ contains only zeros in $\mathtt{first}$, and discard them in the following steps. For every discarded suffix, this strategy saves a $O(m)$ backward search in the bidirectional index.

\item The four-sided range reporting data structure could use the reversed prefixes $\REV{T[1..p_i]}$ rather than the reversed prefixes $\REV{T[1..p_{i}-1]}$ for all $i \in [1..z]$. This would allow checking, at position $k$ in $P$, whether $P[1..k-1]$ ends at a position $j$ of $T$ such that $j$ is the last position of a factor, and such that $T[j+1]=P[k]$. To implement an ever more stringent filter, one could store an additional four-sided range reporting data structure that uses the reversed prefixes $\REV{T[1..p_{i}-1+h]}$ for $h>1$, reverting to the four-sided data structure for $h=1$ when $k>m-h$.

\item In the bidirectional index, we could quit the synchronized backward search for $P[1..k-1]$ in $\BWT_{T\#}$ as soon as we find a suffix $P[i..k-1]$ which is not right-maximal, and which is not followed by $P[k]$ in $T\#$, or which does not contain a factor as a suffix. Moreover, we could quit the backward search as soon as we detect that $P[i..k-1]$ is right-maximal, but it is neither a factor nor the suffix of a factor, and it does not contain a factor as a suffix. Such tests can be implemented e.g. using variations of function $\INTERVALFUNCTION$, and by replacing the interval of $P[1..k-1]$ with the $\sigma+1$ subintervals of $P[1..k-1] \cdot a$ for all $a \in [0..\sigma]$, as described in \cite{belazzougui2014linear}. Every backward step would have a $O(\sigma)$ overhead in this case.

\item In the bidirectional index, we could speed up the backward search that we have to perform for every $P[1..k-1]$, by precomputing a table of intervals in $\BWT_{T\#}$ and $\BWT_{\REV{T}\#}$ for all strings of length $h$. This would take at most $4\sigma^{h}\log{n}$ bits of additional space, or at most $z(h\log{\sigma}+4\log{n})$ bits if we just store the $h$-mers that suffix a factor.
\end{enumerate}

Due to lack of space, in the main paper we study just the effects of the first two optimizations.

\end{document}